\documentclass[10pt,twocolumn,letterpaper]{article}

 \usepackage{cvpr}              % camera-ready version
\usepackage{cuted} %< for strip environment

\usepackage{caption}

\usepackage{amsmath,amssymb}
\usepackage{graphicx}
\usepackage{booktabs}
\usepackage{multirow}
\usepackage{algorithm}
\usepackage{algpseudocode}
\usepackage{float}
\usepackage{xcolor}

\definecolor{cvprblue}{rgb}{0.21,0.49,0.74}
\definecolor{tbdred}{rgb}{0.8,0.1,0.1}

\definecolor{cvprblue}{rgb}{0.21,0.49,0.74}
\usepackage[pagebackref,breaklinks,colorlinks,allcolors=cvprblue]{hyperref}

\def\confName{CVPR}
\def\confYear{2027}

\title{FM-Mamba: An Efficient Native 2D Selective State Space Model with Stability-Aware Discretization}

\author{
Shengyuan Hu\\
School of Computer Science, Beijing Institute of Technology\\
Beijing, China\\
{\tt\small shengyuan.hu@bit.edu.cn}
\and
Yiyong Sun\thanks{Corresponding author.}\\
School of Aerospace Engineering, Beijing Institute of Technology\\
Beijing, China\\
{\tt\small sunyy@bit.edu.cn}
\and
Weihang Shan\\
School of Aerospace Engineering, Beijing Institute of Technology\\
Beijing, China\\
{\tt\small bitswh@bit.edu.cn}
\and
Yongzhe Chang\\
Shenzhen International Graduate School, Tsinghua University\\
Shenzhen, China\\
{\tt\small changyongzhe@sz.tsinghua.edu.cn}
}

\begin{document}
\maketitle

\begin{abstract}
When applied to two-dimensional features such as images, Mamba-based methods
either flatten the feature map into 1D sequences and fuse multiple scans,
which breaks spatial adjacency and multiplies computation, or recurse
directly on the 2D grid but rescale the recurrence with a fixed
input-independent factor and aggregate only the predecessors on a single
anti-diagonal, leaving the discrete step unconstrained by 2D stability
theory.
We propose \textbf{FM-Mamba}, a native 2D selective state space model built
on the general Fornasini--Marchesini (FM) model.
Its coupled recurrence evolves along both spatial axes; cross-axis coupling
induces a discrete diagonal-predecessor term $\tilde A_0$, and two
independent input-dependent step sizes $\Delta_1(i,j)$, $\Delta_2(i,j)$
govern propagation along the two axes, so that the coefficients adapt per
position and per direction.
A normalized ratio $r_k=a_{0,k}/(a_{1,k}a_{2,k})$ yields a closed-form
piecewise bound on the two step sizes; when a configuration falls outside
the stable interval the steps are projected onto this bound, and the
criterion is re-verified per position after training, which suppresses
systematic local amplification without global damping.
The recurrence is executed by a dual-diagonal wavefront scan: each update
gathers information from two anti-diagonals, and the input enters the current
state through a single projection, so that neither training nor inference
flattens the feature map.
On COCO 2017 multi-label classification, replacing the Mamba2D scan core
with FM-Mamba reduces training time by up to $52.8\%$, and peak memory and
training energy by $42.1\%$ and $53.5\%$, with accuracy within $0.12$ percentage points of it; using FM-Mamba
in all stages cuts parameters by $11.42\%$ and raises throughput from $53.76$
to $86.74$\,FPS, and to $385.39$\,FPS with the optimized PSD implementation.
Code is available at \url{https://github.com/BitSsm/FM-Mamba-Runtime}.
\end{abstract}

\section{Introduction}
\label{sec:intro}

State space models (SSMs) scale linearly with sequence length where Transformers scale quadratically, which makes them preferable for long sequences. S4 \cite{gu2022efficiently} showed that a continuous-time SSM initialized with HiPPO \cite{gu2020hippo} captures long-range dependencies effectively, and Mamba \cite{gu2023mamba} makes the discretization step, the input matrix and the output matrix input-dependent, thus realizing attention-like content-adaptive information filtering.

When applied to two-dimensional features such as images, however, this recurrence faces a structural mismatch: the signal is defined on a spatial grid, whereas the recurrence acts on a 1D sequence. Flattening the grid into sequences, as vision Mamba methods do \cite{zhu2024vision,liu2024vmamba,zhang20242dmamba,li2024mamband}, therefore carries three consequences, none of which a scan order can remove entirely. (i) Spatial adjacency is destroyed: under raster order a vertical neighbour lies $W$ steps away while a horizontal one lies $1$ step away, so the receptive field is compressed along one axis and the two directions are treated asymmetrically. (ii) Multipath propagation is lost: on the grid a signal at $(i',j')$ reaches $(i,j)$ along $\binom{(i-i')+(j-j')}{i-i'}$ monotone paths, whereas a chain provides exactly one. (iii) Multi-directional scanning repeats the recurrence several times per layer \cite{zhu2024rethinking,zhu2024vision,liu2024vmamba}, multiplying the computation and accumulating 2D--1D--2D conversion overhead.

Native 2D recurrences \cite{gu2022s4nd,baron2024twodim,zhang20242dmamba,baty2026mamba2d,wang2024v2m,yu20262d,li2024mamband} remove the flattening but inherit two separate gaps: those that keep the diagonal predecessor provide no criterion constraining the discrete step, while those that constrain the step give up the diagonal predecessor and bound the recurrence by a fixed global scaling of the two directional contributions \cite{baty2026mamba2d}. Scan design closes neither gap, since both concern the coefficient structure rather than the traversal order. The two gaps are independent: the diagonal predecessor adds a propagation path that criteria derived for one-predecessor recurrences do not cover, whereas global scaling applies the same damping at every position and does not extend to a recurrence that keeps the diagonal term.

Existing remedies split along the same line. Scan-design methods \cite{zhao2022rethinking,xiao2025boosting,xie2024quadmamba,li2025damamba} keep the 1D recurrence and treat the conversion as a traversal problem, so the mismatch is narrowed only within what an ordering can recover: different orders differ little in accuracy and none dominates \cite{zhu2024rethinking}, while every additional direction is paid for by further scans and further 2D--1D--2D conversion. Native 2D recurrences \cite{zhang20242dmamba,baty2026mamba2d} settle the coefficient problem in one of two ways: stability is taken from the scale of the coefficients, or it is enforced by a construction that cannot be switched off.

This paper builds \textbf{FM-Mamba} on the single-input form of the general FM model \cite{kurek1985general}: 2D continuous dynamics discretized by per-axis ZOH with an independent step along each direction, in-place injection of the input $x(i,j)$ into the current state $h(i,j)$ through a single projection, and a dual-diagonal wavefront parallel scan, with no flattening at any stage. The main contributions are as follows:
\textbf{1. Dual-diagonal native 2D recurrence.} Per-axis ZOH discretization of the 2D continuous single-input model makes the diagonal-predecessor term $\tilde A_0$ arise from cross-axis coupling, so each update aggregates three predecessors on two anti-diagonals and a third propagation path opens between the two 1D chains of the grid; the gain is in solution and classification quality rather than in speed.
\textbf{2. Single-drive in-place injection and anisotropic selection.} The input enters the current state in place through one projection, one input matrix fewer than a horizontal/vertical dual drive, and two input-dependent steps parameterize the horizontal and vertical transitions separately, so the relative weight between the two directions is decided per position by the content.
\textbf{3. Stability-aware discretization.} A closed-form per-position criterion characterizes, through the normalized ratio $r_k=a_{0,k}/(a_{1,k}a_{2,k})$, the piecewise requirement that the frozen-coefficient 2D stability criterion imposes on the discrete step; configurations outside the stable interval are handled by a closed-form upper-bound projection of the step and a post-training per-position verification.

\section{Related Work} \label{sec:related}
A 1D state space model maps an input $x$ to an output $y$ through a hidden state $h\in\mathbb{R}^n$; its continuous form discretized by zero-order hold (ZOH) gives
\begin{equation}
\left\{
\begin{aligned}
\dot h(t)&=A h(t)+B x(t)\\
y(t)&=C h(t)+D x(t)
\end{aligned}
\right.
\Rightarrow
\left\{
\begin{aligned}
h_k&= {A_k} h_{k-1}+ {B_k} x_k\\
y_k&=C_k h_k+D x_k
\end{aligned}
\right.
\label{eq:cont_ssm}
\end{equation}
where ${A_k}$ comes from the input-independent continuous parameter $A$, while the step size and the input/readout vectors ${B}_k,C_k$ are generated per position from the input \cite{gu2023mamba}: $\Delta_k=\mathrm{Softplus}(b_\Delta+W_\Delta x_k)$, $A_k=e^{\Delta_k A}$, $B_k=(\Delta_k A)^{-1}(e^{\Delta_k A}-I)\,\Delta_k B$; implementations often use the first-order approximation $B_k\approx\Delta_k B_t$, where $B_t=W_Bx_k$; $C_k=W_Cx_k$.
Selection makes the parameters vary with position and the kernel no longer fixed, so the recurrence is computed by a parallel prefix scan of $O(L)$ work, which reads as time-varying filtering: the step sets forgetting through $A_k$ and writing through $B_k$.

\textbf{Flattening family.} Vision Mamba methods \cite{zhu2024vision,liu2024vmamba,li2025damamba,huang2024localmamba,yang2024plainmamba,xiao2025boosting,xie2024quadmamba} keep this 1D selective recurrence and address the 2D--1D conversion along three lines: scan order, with raster/Zigzag \cite{zhao2022rethinking}, Hilbert fractal \cite{xiao2025boosting}, quadtree \cite{xie2024quadmamba} and dynamic adaptive \cite{li2025damamba} paths; 2D-aware serialization with a hardware-aligned operator, as in 2DMamba \cite{zhang20242dmamba}, whose period-layered scan order narrows the mismatch on whole-slide classification while its state update remains 1D; and hybrid or task-specific backbones, where MambaVision \cite{hatamizadeh2025mambavision} combines Mamba with attention, MambaX \cite{li2026mambax} learns the coefficients of the state equation dynamically for super-resolution, and MTMamba++ \cite{lin2025mtmambaplus} models intra-task and cross-task dependencies for dense scene understanding. Comparative evidence \cite{zhu2024rethinking} reports that scan directions and their combinations differ little in accuracy, that no order dominates, and that multi-directional fusion raises the cost substantially. In all of these the recurrence stays 1D, the grid enters only through the order in which the chain is read, so the coefficient structure never becomes two-dimensional and any 2D interaction is at best approximated by the input ordering.

\textbf{Native 2D recurrence.} S4ND \cite{gu2022s4nd} applies multi-dimensional SSMs to images and video separably, and \cite{baron2024twodim} embeds a Roesser two-substate layer into Transformer blocks as a plug-in, with input-independent parameters. Mamba2D \cite{baty2026mamba2d} keeps a single hidden state and performs an additive joint recurrence equivalent to the FM-I structure; it maintains stability by scaling the two directional contributions with a fixed $1/2$ weight, injects the input through horizontal/vertical dual projections, and demonstrates native 2D recurrence at large vision scale. V2M \cite{wang2024v2m} builds horizontal/vertical substates on the Roesser model and approximates the 2D recurrence by row-wise and column-wise 1D SSMs for hardware efficiency; 2D-CrossScan Mamba \cite{yu20262d} aggregates hidden states from a 2D neighbourhood, scans the four corners with trainable fusing weights, and shares one input-dependent step between the two directions; \cite{behrouz2024chimera} applies parallel selective 2D scanning to multivariate time series, while Mamba-ND \cite{li2024mamband} alternates row-major orders across layers as a serialization transformation. Native 2D recurrence is thus computationally feasible, yet it carries no diagonal-predecessor term $\tilde A_0$ and is unconstrained by 2D stability theory: the expressiveness of that term and its stability tractability never come together in these models.

Two-dimensional discrete systems theory constrains the coefficient field of a recurrence rather than its layout on the grid, and provides the framework for constructing and constraining a selective SSM on a 2D grid: the Fornasini--Marchesini first model (FM-I) \cite{fornasini1976state} describes the state evolution with a single local state and no explicit diagonal predecessor, the Roesser model \cite{roesser1975discrete} introduces horizontal and vertical substates, and the general FM state space model \cite{kurek1985general,wu2015filtering,yang2019two}  adds the diagonal-predecessor and direct input terms, with 2D asymptotic stability criteria \cite{bachelier2023connections,kaczorek1985two}.
This paper adopts the general FM-type recurrence: its diagonal-predecessor term arises from cross-axis coupling in the per-axis ZOH discretization, while a single hidden state avoids the extra dimension of the Roesser model. The discrete state space model of the general FM model is
\begin{equation}
\begin{aligned}
h(i,j)&=\bar A_1 h(i{-}1,j)+\bar A_2 h(i,j{-}1)\\
 &+\bar A_0 h(i{-}1,j{-}1)+\cdots,
\label{eq:fm_general}
\end{aligned}
\end{equation}
Let $\mathbb U=\{z\in\mathbb C:\,|z|<1\}$ denote the open unit disc and $\overline{\mathbb U}^2=\{(z_1,z_2):\,|z_1|\le1,\ |z_2|\le1\}$ the closed unit bidisc. The \emph{2D characteristic polynomial}  is defined as
\begin{equation}
\Delta(z_1,z_2)\triangleq\det\!\left(I-z_1\bar A_1-z_2\bar A_2-z_1z_2\bar A_0\right),
\label{eq:char_poly}
\end{equation}
and it is asymptotically stable if and only if $\Delta(z_1,z_2)$ has no zero in the closed unit bidisc \cite{kurek1985general,bachelier2023connections}:
\begin{equation}
\Delta(z_1,z_2)\neq0,\qquad\forall\,(z_1,z_2)\in\overline{\mathbb U}^2.
\label{eq:stab_bidisc}
\end{equation}
The boundary of the closed disc cannot be omitted: a zero on $|z_1|=1$ or $|z_2|=1$ corresponds to the critical case, in which the recurrence no longer decays. Equation~\eqref{eq:stab_bidisc} is a global criterion on a two-variable domain and cannot be reduced to a finite number of one-dimensional tests for general matrices \cite{bachelier2023connections}; only when $\bar A_1,\bar A_2,\bar A_0$ are simultaneously diagonal does the system decouple into $n$ scalar 2D systems.
Specifically, taking the diagonal discrete form ($\bar A_1=\mathrm{diag}(\bar a_{1,1},\ldots,\bar a_{1,n})$, $\bar A_2=\mathrm{diag}(\bar a_{2,1},\ldots,\bar a_{2,n})$, $\bar A_0=\mathrm{diag}(\bar a_{0,1},\ldots,\bar a_{0,n})$), the scalar recurrence $h_k(i{+}1,j{+}1)=\bar a_{1,k}h_k(i,j{+}1)+\bar a_{2,k}h_k(i{+}1,j)+\bar a_{0,k}h_k(i,j)$ of the $k$-th channel is asymptotically stable if and only if $|\bar a_{2,k}|<1$ together with
\begin{equation}
|\bar a_{1,k}+\bar a_{0,k}|<1-\bar a_{2,k}\ \text{and}\ |\bar a_{1,k}-\bar a_{0,k}|<1+\bar a_{2,k},
\label{eq:scalar_stab}
\end{equation}
This criterion applies to the \emph{FM-type scalar recurrence} used in this paper, and follows from substituting the channel's characteristic polynomial $1-z_1\bar a_{1,k}-z_2\bar a_{2,k}-z_1z_2\bar a_{0,k}$ into~\eqref{eq:stab_bidisc}; it depends only on the 2D recurrence relation among the discrete coefficients, not on their origin.

\section{Method}
\label{sec:method}

\subsection{Continuous Model and Aperiodic Discretization}
\label{sec:disc}

\paragraph{Notation.}
The continuous diagonal matrices are $A_1,A_2,A_0$, with $a_{1,k},a_{2,k},a_{0,k}$ their $k$-th diagonal elements; two sets of discrete coefficients are distinguished, the fixed-step coefficients (all positions sharing the same step) $\bar A_1,\bar A_2,\bar A_0,\bar B,\bar C$ with elements $\bar a_{1,k},\bar a_{2,k},\bar a_{0,k}$, and the input-dependent selective coefficients $\tilde A_1,\tilde A_2,\tilde A_0,\tilde B,\tilde C$ with elements $\tilde a_{1,k},\tilde a_{2,k},\tilde a_{0,k}$.
The input projection $W_{B_0}$ and the output projection $W_C$ map the feature vector $X(i,j)$ at position $(i,j)$ to an $n$-dimensional injection vector and an $n$-dimensional readout vector, $n$ being the state dimension, and the skip term $D$ is a per-channel scalar as in Mamba. With $1\le k\le n$ the diagonal-element index, the equations are written for a single channel: $x$ and $y$ are its scalar input and output at position $(i,j)$, and $h\in\mathbb{R}^n$ is its state, each channel holding such a state and recursing in parallel.
With the continuous single-input 2D FM system as the core, its dynamic model for one channel at one position is written as
\begin{equation}
\left\{\!\begin{aligned}
\frac{\partial^2 h}{\partial t_1\partial t_2}
&=A_1\frac{\partial h}{\partial t_1}+A_2\frac{\partial h}{\partial t_2}+A_0h+B_0x,\\
y(t_1,t_2)&=C^{\top}h(t_1,t_2)+D\,x(t_1,t_2),
\end{aligned}\right.
\label{eq:cont_si}
\end{equation}
where $B_0,C $ are the injection and readout vectors and $X(i,j)$ collects the inputs of all channels at that position. To handle a discrete image grid, we apply an input-dependent \emph{aperiodic} zero-order hold (ZOH) discretization along each direction (full derivation in Appendix~A); eliminating the intermediate state gives the \emph{in-place injection} selective FM-Mamba recurrence
\begin{equation}
\left\{\!\begin{aligned}
h(i,j)&=\tilde A_1(i,j)h(i{-}1,j)+\tilde A_2(i,j)h(i,j{-}1)\\
&\quad+\tilde A_0(i,j)h(i{-}1,j{-}1)+\tilde B(i,j)x(i,j),\\
y(i,j)&=\tilde C(i,j)^{\top}h(i,j)+D\,x(i,j),
\end{aligned}\right.
\label{eq:sel_si}
\end{equation}
where the three matrices $\tilde A_0$, $\tilde A_1$, and $\tilde A_2$  are $n\times n$ diagonal calculated as
$\tilde A_1(i,j)=e^{A_2\Delta_1}$,  $\tilde A_2(i,j)=e^{A_1\Delta_2}$,
$\tilde A_0(i,j)=G(A_2,\Delta_1)G(A_1,\Delta_2)(A_1A_2+A_0)
 -e^{A_1\Delta_2}e^{A_2\Delta_1}$,
$\tilde B(i,j),\tilde C(i,j) $ are the injection and readout vectors generated from the input features at that position.
The discrete coefficients at each position follow directly from the two input-dependent steps and the input vector
\begin{equation}
\begin{aligned}
&\Delta_1(i,j)=\mathrm{Softplus}\bigl(b_{\Delta_1}+w_{\Delta_1}^{\top}X(i,j)\bigr),\\
&\Delta_2(i,j)=\mathrm{Softplus}\bigl(b_{\Delta_2}+w_{\Delta_2}^{\top}X(i,j)\bigr),\\
&\tilde B(i,j)=G(A_2,\Delta_1)\,G(A_1,\Delta_2)\,W_{B_0}X(i,j),\\
&\tilde C(i,j)=W_CX(i,j),
\end{aligned}
\label{eq:sel_proj}
\end{equation}
where the integral matrix $G(A,\Delta)\triangleq\int_0^{\Delta}e^{As}\,ds$ describes the accumulated injection over the anisotropic step. The two steps enter the exponents crosswise: $\Delta_1$ multiplies $A_2$ and $\Delta_2$ multiplies $A_1$ (Appendix~A). The same ZOH formulas generate the diagonal predecessor as well: $\tilde A_0$ may remain nonzero even when $A_0$ is zero, because the cross-axis product $A_1A_2$ enters its expression. The parameters $A_0,A_1,A_2$ are fixed in advance and are not trained; the skip term $D$ is trained but stays input-independent, and the learnable parameter set is
\begin{equation}
\begin{aligned}
\Theta={}&\bigl\{W_{B_0},\ W_C,\ b_{\Delta_1},b_{\Delta_2},\ w_{\Delta_1},w_{\Delta_2},\ D\bigr\},
\end{aligned}
\label{eq:params}
\end{equation}
so the model keeps one input and one output projection with few learned parameters. Following Mamba, the implementation uses the first-order approximation $G(A,\Delta)\approx\Delta I$, which reduces the input injection to the pointwise product used in Mamba:
\begin{equation}
\tilde B(i,j)\approx\Delta_1(i,j)\Delta_2(i,j)\,W_{B_0}X(i,j).
\label{eq:first_order_B}
\end{equation}
This approximation acts only on the input term $\tilde B$: the state coefficients keep the exact matrix exponentials and integrals of Eq.~\eqref{eq:sel_proj}, and since $A_1,A_2,A_0$ are diagonal only their closed-form values are needed per position, with no numerical integration. The two independent steps thus adjust the horizontal and vertical propagation separately, the anisotropic content-adaptive propagation used throughout.
The two steps $\Delta_1(i,j)$ and $\Delta_2(i,j)$ are generated independently and need not coincide, although classical 2D discrete systems theory assumes a common sampling period.

\subsection{Stability Analysis and Sampling-Period Constraint}
\label{sec:stab}

The discrete coefficients are produced from the continuous system by ZOH discretization and are input-dependent, so their stability must be examined in the discrete domain. Since $a_{1,k},a_{2,k}<0$ originate from the HiPPO--LegS diagonal initialization~\cite{gu2020hippo,gu2023mamba} and the continuous matrices are diagonal, the coefficients stay diagonal after discretization and the state components stay decoupled.
Freezing at each position the discrete coefficients generated by $(\Delta_1(i,j),\Delta_2(i,j))$, and applying the scalar criterion~\eqref{eq:scalar_stab} to the $k$-th component, gives the per-position criterion
\begin{equation}
\left\{\!\begin{aligned}
|\tilde a_{1,k}(i,j)+\tilde a_{0,k}(i,j)|&<1-\tilde a_{2,k}(i,j),\\
|\tilde a_{1,k}(i,j)-\tilde a_{0,k}(i,j)|&<1+\tilde a_{2,k}(i,j),
\end{aligned}\right.
\label{eq:perpos_stab}
\end{equation}
where $\tilde a_{2,k}(i,j)\in(0,1)$ automatically satisfies the premise $|\tilde a_{2,k}|<1$. Both inequalities are the bidisc test~\eqref{eq:stab_bidisc} evaluated at one point of the coefficient field, requiring that $1-z_1\tilde a_{1,k}-z_2\tilde a_{2,k}-z_1z_2\tilde a_{0,k}$ has no zero on $|z_1|=|z_2|=1$ for the coefficients frozen there.
The criterion characterizes the stationary recurrence obtained by freezing the coefficients over the grid and serves as a conservative design condition, ruling out sustained local amplification; its scope is detailed in Appendix~B.

The relation between $A_0$ and the product $A_1A_2$ is characterized by a normalized ratio:
\begin{equation}
r_k\triangleq \frac{a_{0,k}}{a_{1,k}a_{2,k}},\quad a_{0,k}=r_k\,a_{1,k}a_{2,k},
\label{eq:r}
\end{equation}
so that any continuous diagonal $A_0$ is characterized uniquely by $r_k$ ($A_0=-A_1A_2$ gives $r_k=-1$, and $A_0=0$ gives $r_k=0$, the prototype without diagonal predecessor). Substituting Eq.~\eqref{eq:r} into the ZOH coefficients of Appendix~A and then into criterion~\eqref{eq:scalar_stab} gives the piecewise boundary in $r_k$: for $-2<r_k<0$ the whole first quadrant $(\Delta_1,\Delta_2)\in(0,+\infty)^2$ is stable for arbitrary positive steps; for $r_k<-2$ the steps must fall into a bounded rectangle; $r_k=-2$ is the critical value, at which the step is theoretically unconstrained. The branch $r_k\ge0$ admits no asymptotically stable choice of positive steps (Appendix~B.3), where $r_k=0$ is critically stable for every step, so the parameterization is restricted to $r_k<0$.

Each position $(i,j)$ corresponds to a pair of steps $\Delta_1,\Delta_2$ shared by all channels and the $n$ state components, so the tightest bound over components gives the scalar step bounds
\begin{equation}
\begin{aligned}
\Delta_{1}^{\max}&=\min_{k}\cfrac{-\ln(1+2/r_k)}{|a_{2,k}|},\\
\Delta_{2}^{\max}&=\min_{k}\cfrac{-\ln(1+2/r_k)}{|a_{1,k}|}.
\end{aligned}
\label{eq:delta_star}
\end{equation}
Components with $r_k\in(-2,0)$ are stable for arbitrary positive steps and take no part in the minimum; the default initialization $A_0=-A_1A_2$ gives $r_k=-1$, so the recurrence is per-position stable without any constraint. For fixed $A_0$ with $r_k<-2$, the step is projected by a forward upper bound at every position:
\begin{equation}
\begin{aligned}
\Delta_1(i,j)&\leftarrow\min\{\Delta_1(i,j),(1-\varepsilon)\,\Delta_1^{\max}\},\\
\Delta_2(i,j)&\leftarrow\min\{\Delta_2(i,j),(1-\varepsilon)\,\Delta_2^{\max}\},
\end{aligned}
\label{eq:clip}
\end{equation}
with a margin $\varepsilon$ reserved, because \eqref{eq:delta_star} gives a \emph{critical} bound: at that value the coefficients lie exactly on the boundary of the stable rectangle, where the strict inequality of the criterion fails (a zero of the characteristic polynomial on the closed bidisc, Eq.~\eqref{eq:stab_bidisc}), and rounding, drift of the activation scale and parameter updates may push them across it. We therefore take $\varepsilon=10\%$ uniformly, so that the coefficients always remain inside the rectangle. The projection is applied in the forward pass before discretization; after training, criterion~\eqref{eq:perpos_stab} is evaluated in closed form as a hard verification, and any violating position tightens the bound and triggers retraining. Because the first-order form of \eqref{eq:first_order_B} deviates from the exact integral, the margin is kept uniformly rather than only where the bound is tight, and the critical value $r_k=-2$ is avoided, since its bound cannot be approached by any finite step.

Mamba2D stabilizes its no-predecessor recurrence ($r_k=0$) by a fixed global scaling: $\tilde a_{1,k}/2+\tilde a_{2,k}/2<1$ at every position satisfies criterion~\eqref{eq:perpos_stab} unconditionally, but the damping applies everywhere and cannot be switched off, and with a diagonal predecessor present it constrains nothing, since for $r_k<-2$ the criterion still fails at large steps and a step bound is required in addition. FM-Mamba keeps $\tilde A_0$ without special treatment: with diagonal coefficients the bidisc criterion~\eqref{eq:stab_bidisc} reduces to the per-position test~\eqref{eq:perpos_stab} and the step bound~\eqref{eq:delta_star}, imposed only for $r_k<-2$ and never for $r_k\in(-2,0)$.

\subsection{Dual-Diagonal Wavefront Scan}
\label{sec:scan}

\begin{figure}[h]
\centering
\includegraphics[width=0.95\linewidth]{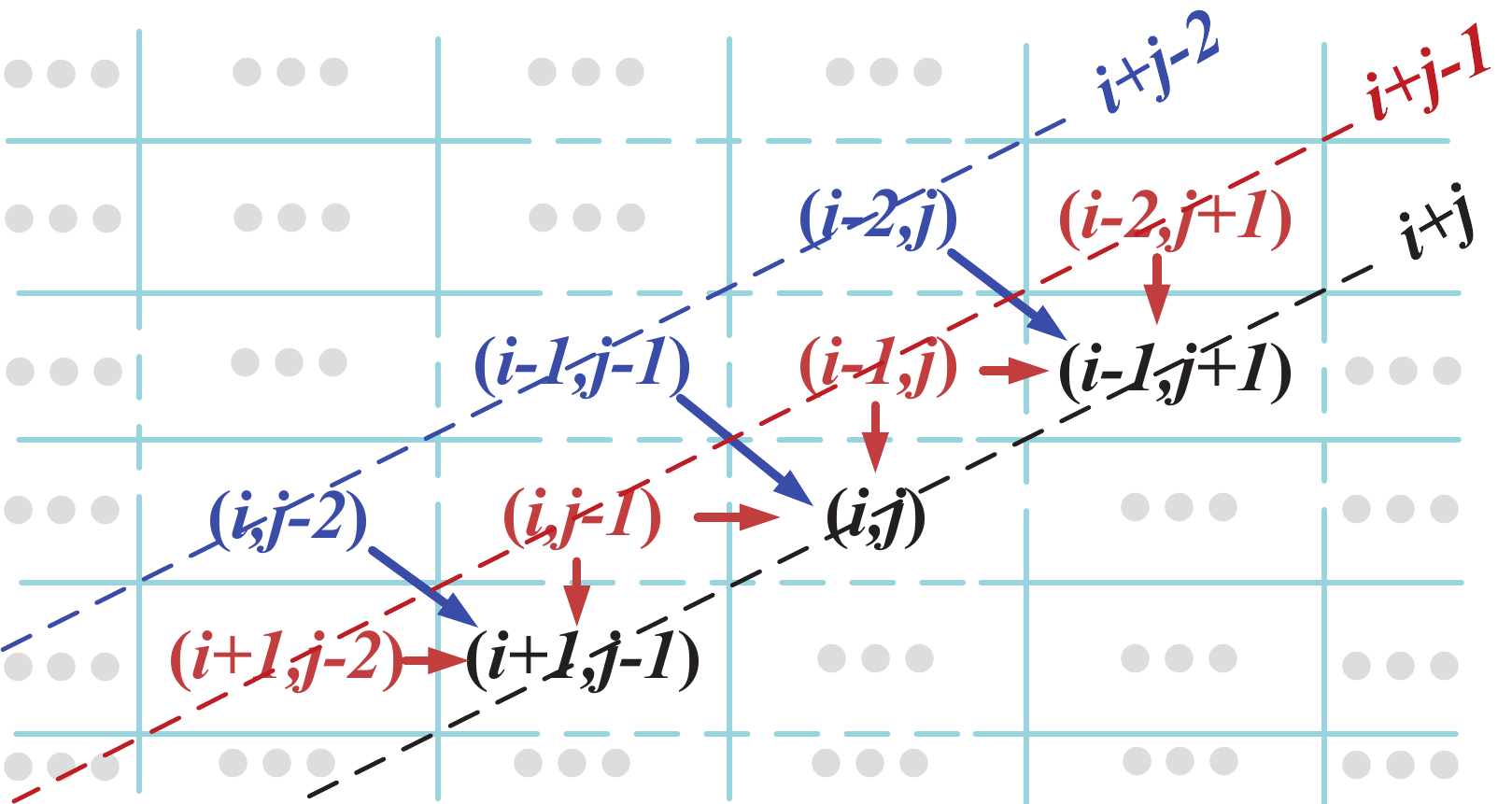}
\caption{Dual-Diagonal Wavefront Scan: black marks the anti-diagonal $i{+}j$ on which $h(i,j)$ resides, red the anti-diagonal $i{+}j{-}1$ read through $\tilde A_1(i,j)h(i{-}1,j)$ and $\tilde A_2(i,j)h(i,j{-}1)$, and blue the anti-diagonal $i{+}j{-}2$ read through $\tilde A_0(i,j)h(i{-}1,j{-}1)$}
\label{fig:wavefront}
\end{figure}

A 2D recurrence cannot be recast as an associative parallel scan the way a 1D prefix scan can: each state $h(i,j)$ depends on three predecessors (Eq.~\eqref{eq:sel_si}, Fig.~\ref{fig:wavefront}), and the accumulated products along different lattice paths are not associative. The dependency is, however, strictly causal along anti-diagonals: $h(i,j)$ depends only on positions with smaller $i{+}j$, so states on the same anti-diagonal $i{+}j=d$ are mutually independent and available in parallel, while the differing diagonal lengths make the load fluctuate. The causality is the same as in single-predecessor 2D cross scans, except that the third predecessor couples two anti-diagonals instead of one. The wavefront therefore runs over $d=0,\dots,H{+}W{-}2$, giving a sequential depth of $H{+}W{-}1$ steps instead of the naive $HW$, with at most $\min(H,W)$ positions in parallel at each step, each holding all channels.
Two properties separate this schedule from the 1D chain and from single-predecessor cross scans (Fig.~\ref{fig:wavefront}):
\begin{itemize}
\item \textbf{Dual-diagonal dependency.} $h(i,j)$ is driven by the two predecessors on the anti-diagonal $i{+}j{-}1$ and by the diagonal predecessor on $i{+}j{-}2$, so information from two anti-diagonals converges into the current state and the multipath propagation of the 2D recurrence is preserved.
\item \textbf{Single-drive in-place injection.} The input enters the current state through one projection, acts at the current position and does not accumulate along the scan, so one input matrix $B$ is saved at the same wavefront depth.
\end{itemize}
Algorithm~\ref{alg:scan} gives the procedure: step $d$ depends only on the two preceding frames $\mathcal{F}_{d-1}$ and $\mathcal{F}_{d-2}$.

\begin{algorithm}[h]
\caption{Dual-Diagonal Wavefront Scan}
\label{alg:scan}
\begin{algorithmic}[1]
\Require Feature map $X$; continuous matrices $A_1,A_2,A_0$; projections $W_{B_0},W_C$; step weights and biases; skip term $D$
\Ensure Output $Y$
\State Compute $\Delta_1,\Delta_2$ and $\tilde C=W_CX$ for all $(i,j)$ in parallel \Comment{one scalar step pair per position}
\State If $r_k<-2$, project the steps onto Eq.~\eqref{eq:delta_star} with the $10\%$ margin \Comment{Eq.~\eqref{eq:clip}, before discretization}
\State Form $\tilde A_1,\tilde A_2,\tilde A_0$ and $\tilde B$ from the projected steps \Comment{Eqs.~\eqref{eq:sel_proj}, \eqref{eq:first_order_B}}
\State Allocate two buffers for the states on anti-diagonals $\mathcal{F}_{d-1},\mathcal{F}_{d-2}$ \Comment{the $\tilde A_0$ term is one wavefront older}
\State Off-grid predecessors are zero: $h(-1,j)=h(i,-1)=h(-1,-1)=0$ \Comment{no boundary branch needed}
\For{$d=0$ \textbf{to} $H+W-2$} \Comment{$H+W-1$ steps, $\min(H,W)$ positions per step}
  \State \textbf{parallel for} all $(i,j)$ with $i+j=d$, over all channels
  \State Read $h(i{-}1,j),h(i,j{-}1)$ from $\mathcal{F}_{d-1}$ and $h(i{-}1,j{-}1)$ from $\mathcal{F}_{d-2}$
  \State $h(i,j)\leftarrow\tilde A_1(i,j)h(i{-}1,j)+\tilde A_2(i,j)h(i,j{-}1)+\tilde A_0(i,j)h(i{-}1,j{-}1)+\tilde B(i,j)x(i,j)$
  \State $y(i,j)\leftarrow\tilde C(i,j)^{\top}h(i,j)+D\,x(i,j)$ \Comment{$h$ and $y$ are written in place}
  \State Store $h(i,j)$ into the older buffer and swap the pointers \Comment{residency $O(\min(H,W))$, independent of $HW$}
\EndFor
\State \Return $Y$
\end{algorithmic}
\end{algorithm}

FM-Mamba performs one state update and one output projection per position, so the total compute is $O(HWn)$; the wavefront depth is $O(H{+}W)$ with a parallelism of $\min(H,W)$, the same as for a single-predecessor 2D recurrence, and double buffering adds only a constant factor of storage. Flattening alternatives need $2$--$4$ scans with FLOPs of $2HWn$--$4HWn$ and sequential depths of $2HW$--$4HW$, and all introduce the 2D--1D--2D conversion, whereas a native 2D recurrence preserves the coupling it is built on.

% =========================================================================
% Experiments section: data presentation and objective comparison only.
% Tables: tab:train_quality, tab:Inference_Comparison; figure: fig:train_curves.
% =========================================================================
\raggedbottom
\section{Experiments}
\label{sec:experiments}

\setlength{\parindent}{1pc}
\clubpenalty=10000
\widowpenalty=10000
\displaywidowpenalty=10000
\setlength{\intextsep}{8pt}

Experiments are conducted on COCO 2017 for multi-label classification: after removing 3 unusable images from train2017, the training and test sets are split at a ratio of 8:2, and validation uses all 5{,}000 images of val2017; the test index is not mounted during training. Each image is labeled by the union of its instance categories, giving an 80-dimensional binary label; no detection head is used. Images are resized to $224\times224$ with random horizontal flipping and perturbation of brightness, contrast and saturation; no augmentation is applied at validation or test time. The baselines are the flattening methods Vision Mamba~\cite{zhu2024vision}, VMamba~\cite{liu2024vmamba} and 2DMamba~\cite{zhang20242dmamba}, plus the native 2D method Mamba2D~\cite{baty2026mamba2d}; all structures adopt the four-stage design of Mamba2D-T in~\cite{baty2026mamba2d} (stage depths $[3,3,9,3]$, channel widths $[64,128,320,512]$), where $M_{\rm2D}$ and $M_{\rm FM}$ denote the local branch modules of~\cite{baty2026mamba2d} and of this paper, $A_{\rm 2D}$ and $A_{\rm FM}$ the attention modules, $A_{\rm T}$ the Transformer attention module, and the numeric prefix the number of stages covered. All methods share the hardware (RTX 5090\,D, 32\,GiB), the BCE loss, the AdamW optimizer, the learning rate and the dropout schedule, with $n=16$.

In the continuous system~\eqref{eq:cont_si}, $a_{1,k}=a_{2,k}=-k$ ($k=1,\ldots,16$); the three conditions differ in the elements $a_{0,k}$ of $A_0$ and in the step constraint they induce (Table~\ref{tab:fm_settings}): conditions~$\mathbf{I}$ and $\mathbf{II}$ keep $r_k\in(-2,0)$ and are stable for arbitrary positive steps according to~\S\ref{sec:stab},  whereas condition~$\mathbf{III}$ has $r_k=-2.5$, leaves that interval, and takes the projection of Eq.~\eqref{eq:clip}.  The conditions are applied at both replacement depths $2M_{\rm FM}2A_{\rm T}$ and $2M_{\rm FM}2A_{\rm FM}$, forming an ablation of the stability configuration; $2M_{\rm FM}2A_{\rm FM}$ has in addition the execution-level variant PSD (Per-Shape Dispatch), whose recurrence and parameters are identical.

\begin{table}[!htb]
\centering
\setlength{\abovecaptionskip}{4pt}
\setlength{\belowcaptionskip}{4pt}
\caption{ The three conditions of $A_0$}
\label{tab:fm_settings}
\small
\setlength{\tabcolsep}{4pt}
\begin{tabular}{@{}cccc@{}}
\toprule
Condition & $a_{0,k}$ & $r_k$ & Step constraint\\
\midrule
$\mathbf{I}$   & $-k$      & $[-1,\,-0.0625]$ & none\\
$\mathbf{II}$  & $-k^2$    & $-1$             & none\\
$\mathbf{III}$ & $-2.5k^2$ & $-2.5$           & $\leq 0.9\ln(5)/16$\\
\bottomrule
\end{tabular}
\end{table}

\subsection{Training Efficiency and Quality}
\label{sec:exp-train}

\begin{figure*}[!t]
\centering
\includegraphics[width=\textwidth]{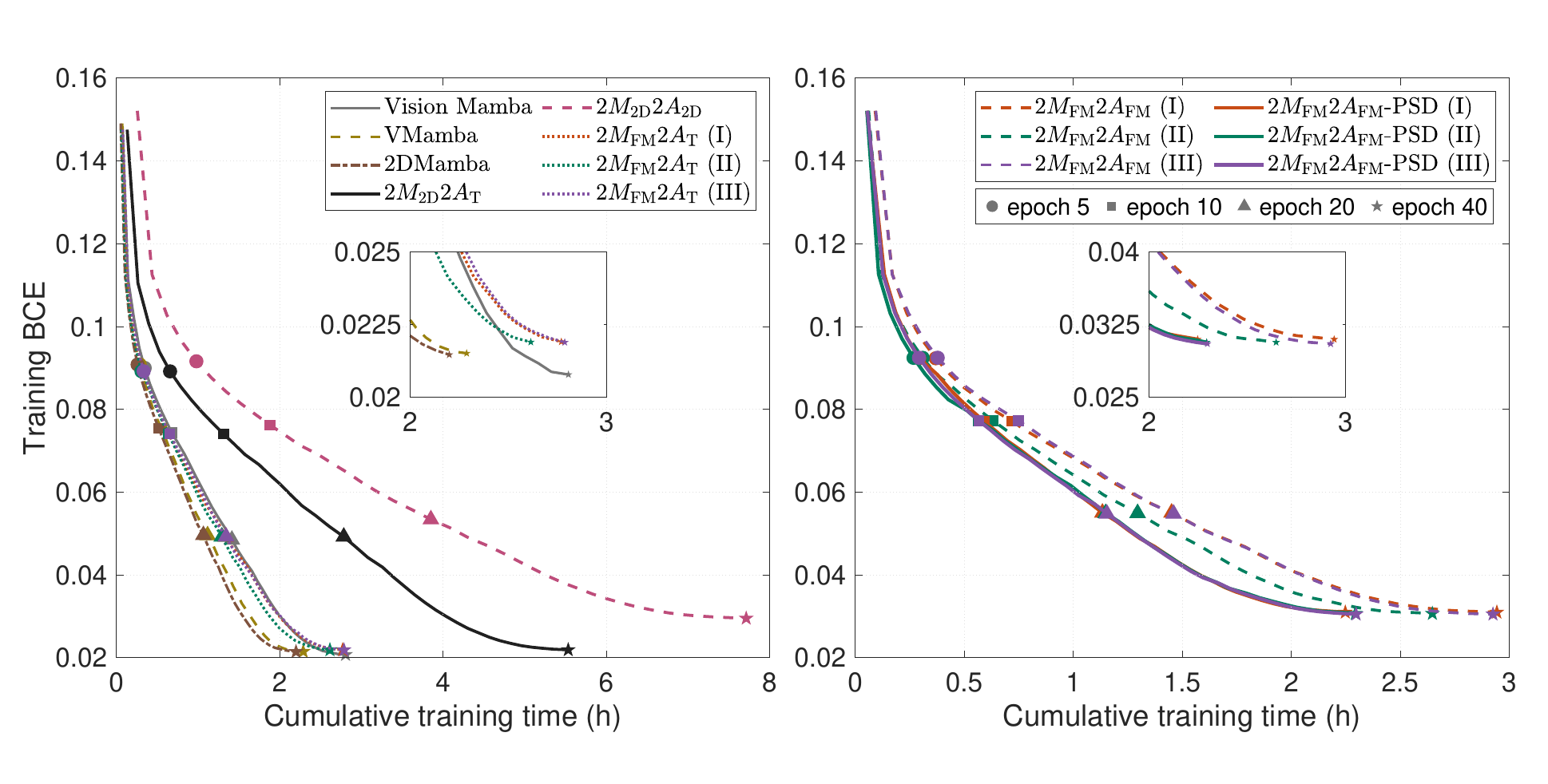}
\caption{Training process comparison. Left: the flattening baselines, $2M_{\rm 2D}2A_{\rm T}$ and the three conditions of $2M_{\rm FM}2A_{\rm T}$; right: $2M_{\rm FM}2A_{\rm FM}$ and its PSD variant in three conditions. Circles, squares, triangles and stars mark epochs 5, 10, 20 and 40, respectively; the horizontal axis is cumulative training time.}
\label{fig:train_curves}
\end{figure*}

Table~\ref{tab:train_quality} reports the training cost and validation quality of the 40-epoch runs for the flattening baselines, the native 2D baselines and our structures at two replacement depths and three conditions, followed by PSD; Fig.~\ref{fig:train_curves} shows the training loss versus elapsed time.

\begin{table*}[!t]
\centering
\caption{Training cost and validation quality}
\label{tab:train_quality}
\footnotesize
\setlength{\tabcolsep}{3pt}
\renewcommand{\arraystretch}{1.00}
\resizebox{\textwidth}{!}{%
\begin{tabular}{@{}lcrrrrrccccc@{}}
\toprule
& & \multicolumn{5}{c}{Training cost} & \multicolumn{5}{c}{Validation quality}\\
\cmidrule(lr){3-7} \cmidrule(l){8-12}
Architecture & Condition & \shortstack{Params\\(M)} & \shortstack{Time/\\epoch (s)} & \shortstack{Total\\(h)} & \shortstack{Peak mem.\\(GiB)} & \shortstack{GPU energy\\(kWh)} & Best epoch & \shortstack{mAP\\(\%)} & \shortstack{micro-F1\\(\%)} & \shortstack{macro-F1\\(\%)} & \shortstack{Val.\\BCE}\\
\midrule
Vision Mamba \cite{zhu2024vision} & -- & 24.720 & 252.53 & 2.806 & 14.63 & 1.545 & 20 & 51.260 & 57.176 & 46.372 & 0.0807 \\
VMamba \cite{liu2024vmamba} & -- & 24.365 & 205.87 & 2.287 & 9.76 & 1.097 & 17 & 51.366 & 56.654 & 44.310 & 0.0782 \\
2DMamba \cite{zhang20242dmamba} & -- & 24.365 & 197.91 & 2.199 & 9.85 & 1.096 & 17 & 51.433 & 56.351 & 44.067 & 0.0781 \\
\midrule
$2M_{\rm 2D}2A_{\rm T}$ \cite{baty2026mamba2d} & -- & 24.638 & 497.98 & 5.533 & 21.99 & 2.892 & 22 & 52.034 & 58.293 & 48.151 & 0.0807 \\
$2M_{\rm 2D}2A_{\rm 2D}$ & -- & 22.063 & 694.18 & 7.713 & 25.03 & 4.081 & 34 & 52.581 & 58.550 & 48.617 & 0.0847 \\
\midrule
$2M_{\rm FM}2A_{\rm T}$ & $\mathbf{I}$ & 24.610 & 249.29 & 2.770 & 12.74 & 1.453 & 22 & 52.060 & 58.297 & 48.045 & 0.0807 \\
$2M_{\rm FM}2A_{\rm T}$ & $\mathbf{II}$ & 24.610 & 235.28 & 2.614 & 12.74 & 1.345 & 22 & 52.078 & 58.307 & 48.224 & 0.0806 \\
$2M_{\rm FM}2A_{\rm T}$ & $\mathbf{III}$ & 24.610 & 250.83 & 2.787 & 13.19 & 1.463 & 22 & 52.111 & 58.155 & 48.184 & 0.0806 \\
\midrule
$2M_{\rm FM}2A_{\rm FM}$ & $\mathbf{I}$ & 21.823 & 264.86 & 2.943 & 13.63 & 1.594 & 27 & 52.903 & 58.335 & 48.145 & 0.0802 \\
$2M_{\rm FM}2A_{\rm FM}$ & $\mathbf{II}$ & 21.823 & 238.24 & 2.647 & 13.63 & 1.415 & 26 & 52.743 & 57.983 & 48.003 & 0.0799 \\
$2M_{\rm FM}2A_{\rm FM}$ & $\mathbf{III}$ & 21.823 & 263.26 & 2.925 & 14.13 & 1.602 & 25 & 52.807 & 58.654 & 48.182 & 0.0786 \\
\midrule
$2M_{\rm FM}2A_{\rm FM}-{\rm PSD}$ & $\mathbf{I}$ & 21.823 & 202.29 & 2.248 & 18.65 & 1.231 & 27 & 52.895 & 58.333 & 48.132 & 0.0802 \\
$2M_{\rm FM}2A_{\rm FM}-{\rm PSD}$ & $\mathbf{II}$ & 21.823 & 206.38 & 2.293 & 18.65 & 1.217 & 26 & 52.737 & 57.914 & 47.916 & 0.0799 \\
$2M_{\rm FM}2A_{\rm FM}-{\rm PSD}$ & $\mathbf{III}$ & 21.823 & 206.74 & 2.297 & 19.41 & 1.277 & 25 & 52.793 & 58.680 & 48.111 & 0.0786 \\
\bottomrule
\end{tabular}}
\par\vspace{3pt}
\begin{minipage}{\textwidth}
\footnotesize
Note:
time per epoch and total time are the mean and sum over 40 epochs, excluding validation, weight saving and initialization; GPU energy is estimated by sampled integration during training.
\end{minipage}
\end{table*}

FM-Mamba lowers training time, peak memory and energy in the same direction. With the parameter count on par with the baseline ($24.610$\,M versus $24.638$\,M), $2M_{\rm FM}2A_{\rm T}$ reduces the 40-epoch training time by $49.6\%$--$52.8\%$, peak memory by $40.0\%$--$42.1\%$ and energy by $49.4\%$--$53.5\%$; its per-epoch time is already of the same order as the three flattening baselines, so two replacement stages approach their speed while keeping the native 2D recurrence. Replacing all four stages lowers the parameter count by a further $11.4\%$ and raises the per-epoch time by only $1.3\%$--$6.2\%$ over the two-stage version, still well below a same-depth replacement by Mamba2D.

The saving is not arithmetic: the reference compute per image is on par between the two routes, and the arithmetic of the scan depends only on the state dimension and the number of positions, so it is of the same order before and after the replacement, whereas the measured per-epoch time differs by about a factor of two. It comes from residency and scheduling: single-drive in-place injection keeps one intermediate tensor from the input projection, and the diagonal coefficients enter as per-position scalars, so peak memory, per-epoch time and energy fall together.

Quality improves as the replacement goes deeper. Replacing the last two stages as well reaches a validation mAP of $52.743\%$--$52.903\%$ with $0.240$\,M fewer parameters and $61.8\%$--$65.7\%$ shorter per-epoch time than its Mamba2D counterpart, whose deeper replacement instead costs $39.4\%$ more per-epoch time and $41.1\%$ more energy for an mAP of $52.581\%$; both routes also swap the attention module ($A_{\rm T}\to A_{\rm FM}$), so the comparison covers the two structures as a whole. All three native 2D configurations ($2M_{\rm 2D}2A_{\rm T}$, $2M_{\rm FM}2A_{\rm T}$, $2M_{\rm FM}2A_{\rm FM}$) exceed the validation accuracy of Vision Mamba, VMamba and 2DMamba, and the three unaccelerated FM conditions of $2M_{\rm FM}2A_{\rm T}$ all exceed the $52.034\%$ of $2M_{\rm 2D}2A_{\rm T}$.

The full-plane wavefront advances one anti-diagonal at a time, so one scan issues many small kernel launches, whereas each 1D baseline runs as a single fused kernel; we therefore add a comparable execution-level optimization to FM-Mamba to keep the throughput comparison fair. The Per-Shape Dispatch (PSD) strategy, without changing the recurrence mathematics or the parameters, dispatches through a static per-shape routing table among three equivalent kernels for the same scan (the full-plane wavefront of \S\ref{sec:scan} and its tiled and ring variants) and reuses captured CUDA Graphs, removing the per-wavefront launch overhead and tapping the inference potential of FM-Mamba. With PSD enabled, the per-epoch training time falls by $13.4\%$--$23.6\%$ relative to the unaccelerated run of the same condition, the validation mAP moves by no more than $0.014$ percentage points and each F1 metric by no more than $0.09$, and the advantage over the flattening baselines and over Mamba2D and 2DMamba is preserved; the price is that CUDA Graph residency raises the training peak memory to $18.65$--$19.41$\,GiB.

\begin{table*}[!t]
\centering
\caption{Test-set classification quality and inference performance}
\label{tab:Inference_Comparison}
\footnotesize
\setlength{\tabcolsep}{3pt}
\renewcommand{\arraystretch}{1.00}
\resizebox{\textwidth}{!}{%
\begin{tabular}{@{}lcrrrrrrrr@{}}
\toprule
Architecture & Condition & \shortstack{mAP\\(\%)} & \shortstack{micro-F1\\(\%)} & \shortstack{macro-F1\\(\%)} & \shortstack{micro-P\\(\%)} & \shortstack{micro-R\\(\%)} & FPS & \shortstack{Compute\\(GFLOPs/img)} & \shortstack{Peak mem.\\(MiB)}\\
\midrule
Vision Mamba \cite{zhu2024vision} & -- & 51.281 & 57.647 & 46.870 & 71.092 & 48.479 & 72.96 & 9.094 & 125.32 \\
VMamba \cite{liu2024vmamba} & -- & 50.748 & 56.938 & 44.257 & 72.947 & 46.692 & 63.71 & 7.864 & 123.85 \\
2DMamba \cite{zhang20242dmamba} & -- & 50.832 & 56.970 & 44.219 & 73.140 & 46.655 & 74.09 & 7.871 & 123.85 \\
\midrule
$2M_{\rm 2D}2A_{\rm T}$ \cite{baty2026mamba2d} & -- & 52.517 & 58.685 & 48.409 & 71.257 & 49.884 & 53.76 & 8.700 & 137.88 \\
$2M_{\rm 2D}2A_{\rm 2D}$ & -- & 52.289 & 59.065 & 48.814 & 70.418 & 50.864 & 64.15 & 7.686 & 126.67 \\
\midrule
$2M_{\rm FM}2A_{\rm T}$ & $\mathbf{I}$ & 52.395 & 58.614 & 48.320 & 71.443 & 49.692 & 64.45 & 8.775 & 126.48 \\
$2M_{\rm FM}2A_{\rm T}$ & $\mathbf{II}$ & 52.399 & 58.660 & 48.382 & 71.307 & 49.823 & 60.79 & 8.775 & 126.48 \\
$2M_{\rm FM}2A_{\rm T}$ & $\mathbf{III}$ & 52.403 & 58.580 & 48.405 & 71.356 & 49.684 & 68.58 & 8.775 & 126.48 \\
\midrule
$2M_{\rm FM}2A_{\rm FM}$ & $\mathbf{I}$ & 52.871 & 58.945 & 48.596 & 71.692 & 50.046 & 86.74 & 7.818 & 114.49 \\
$2M_{\rm FM}2A_{\rm FM}$ & $\mathbf{II}$ & 52.696 & 58.585 & 47.988 & 72.659 & 49.078 & 84.91 & 7.818 & 114.49 \\
$2M_{\rm FM}2A_{\rm FM}$ & $\mathbf{III}$ & 52.693 & 58.792 & 48.106 & 71.351 & 49.992 & 83.11 & 7.818 & 114.49 \\
\midrule
$2M_{\rm FM}2A_{\rm FM}-{\rm PSD}$ & $\mathbf{I}$ & 52.873 & 58.960 & 48.602 & 71.698 & 50.065 & 365.06 & 7.818 & 468.17 \\
$2M_{\rm FM}2A_{\rm FM}-{\rm PSD}$ & $\mathbf{II}$ & 52.694 & 58.620 & 48.033 & 72.705 & 49.107 & 385.39 & 7.818 & 468.17 \\
$2M_{\rm FM}2A_{\rm FM}-{\rm PSD}$ & $\mathbf{III}$ & 52.683 & 58.803 & 48.184 & 71.379 & 49.995 & 384.25 & 7.818 & 468.17 \\
\bottomrule
\end{tabular}}
\par\vspace{3pt}
\begin{minipage}{\textwidth}
\footnotesize
Note:
FPS is the average throughput measured over 3 rounds after warm-up, excluding data loading; peak memory excludes the graph-capture phase.
Compute includes convolution, linear projection, attention matrix multiplication and scan operations.
\end{minipage}
\end{table*}

\subsection{Inference Quality and Efficiency}
\label{sec:exp-infer}

Table~\ref{tab:Inference_Comparison} evaluates classification quality on the held-out test set using the best validation weights of each configuration, and measures inference throughput on the same hardware.

The three conditions of $2M_{\rm FM}2A_{\rm FM}$ attain a test mAP of $52.693\%$--$52.871\%$, with the PSD variant of condition~$\mathbf{I}$ reaching $52.873\%$, the highest in Table~\ref{tab:Inference_Comparison}; enabling PSD changes it by no more than $0.010$ percentage points, so the acceleration does not alter the classification results. Replacing only the first two stages, $2M_{\rm FM}2A_{\rm T}$ reaches $52.395\%$--$52.403\%$, within $0.12$ percentage points of the baseline $2M_{\rm 2D}2A_{\rm T}$ ($52.517\%$), with a clearly higher inference throughput ($60.79$--$68.58$\,FPS versus $53.76$\,FPS). The full replacement $2M_{\rm FM}2A_{\rm FM}$ stands $1.41$--$1.59$ percentage points above the strongest flattening baseline Vision Mamba ($51.281\%$) and $1.95$--$2.12$ above VMamba.

The throughput of $2M_{\rm FM}2A_{\rm FM}$ is $83.11$--$86.74$\,FPS, $54.6\%$--$61.3\%$ above that of $2M_{\rm 2D}2A_{\rm T}$ and above the fastest flattening baseline 2DMamba; excluding the PSD variants, it is the only configuration reaching an mAP above $52.5\%$ and $80$\,FPS at once, i.e., 2D accuracy and 1D-level throughput together. Its reference compute of $7.818$\,GFLOPs/img is below every other method except $2M_{\rm 2D}2A_{\rm 2D}$, and its steady-state peak memory of $114.49$\,MiB is the lowest; PSD pays $4.09\times$ the peak memory to raise the throughput to $365.06$--$385.39$\,FPS, $4.2\times$--$4.6\times$ the unaccelerated runs of the same conditions.

\subsection{Stability Constraint: Ablation and Comparison with Global Scaling}
\label{sec:exp-stab}

Table~\ref{tab:train_quality} validates the measures of~\S\ref{sec:stab}. Conditions~$\mathbf{I}$ and~$\mathbf{II}$ hold for arbitrary steps: over 40 epochs neither diverges nor triggers a projection. Condition~$\mathbf{III}$ takes the closed-form bound~\eqref{eq:delta_star} with the projection~\eqref{eq:clip} ($10\%$ margin) and the post-training per-position hard verification, and pays $6.6\%$--$10.5\%$ per-epoch time and $0.45$--$0.50$\,GiB peak memory relative to condition~$\mathbf{II}$, because the out-of-range case keeps the clipping mask in the forward pass and recomputes it in backward, while the projected step adds one scalar operation per position. In quality the three conditions span no more than $0.160$ percentage points of validation mAP, show no consistent ordering, and the constrained condition~$\mathbf{III}$ is not the worst.

Mamba2D carries no discrete stability criterion and suppresses the sum of the two directional contributions with a fixed $1/2$ weight at every position and channel, which satisfies criterion~\eqref{eq:scalar_stab} at $\tilde a_{0,k}=0$ unconditionally (\S\ref{sec:stab}): it bounds no step. The construction here constrains the step only when the ratio leaves the interval, and costs nothing otherwise. Tables~\ref{tab:train_quality} and~\ref{tab:Inference_Comparison} show the per-epoch time, energy and peak memory of $2M_{\rm 2D}2A_{\rm T}$ above those of the three FM conditions, so the explicit construction adds no structural overhead; the overall difference between the two structures still stems mainly from the scan kernel and the injection scheme (\S\ref{sec:exp-train}).

\section{Conclusion}
\label{sec:conclusion}

This paper proposes FM-Mamba, a native 2D selective state space model built on the general Fornasini--Marchesini framework. Instead of flattening an image into a 1D sequence, it keeps one $n$-dimensional hidden state per channel on the 2D grid, without horizontal/vertical substates, and evolves it along both axes through a coupled recurrence, so spatial adjacency and multipath propagation are preserved with no conversion overhead. Per-axis ZOH discretization makes the diagonal-predecessor term $\tilde A_0$ arise from cross-axis coupling, two independent input-dependent steps realize selection along the two directions, and stability is handled explicitly through the normalized ratio $r_k$, which yields a forward bound on the sampling period and a post-training per-position closed-form verification. On COCO image-level multi-label classification, the parameter count falls while the validation mAP rises and the test mAP stays within $0.12$ percentage points of the Mamba2D counterpart; training takes at most half its time and energy, with peak memory more than $40\%$ lower. Future work covers efficient inference for the anisotropic recurrence, higher-dimensional data, and dense prediction.

\section*{LLM Usage Statement}
A large language model was used only for three auxiliary tasks in this manuscript: grammatical checking and copy-editing, \LaTeX{} layout and notation consistency, and code checking. It took no part in the research topic, the method or experimental design, or the derivation of scientific conclusions; all claims, derivations and results were produced by the authors, who take responsibility for the correctness of the paper.

{
    \small
    \bibliographystyle{ieeenat_fullname}
    \bibliography{Ref20261016}
}

% Supplementary material, following the author-kit: it is appended to this same
% document.  Uncomment to produce the "with supplementary" PDF, then delete the
% supplementary pages before submission.
\appendix
% Long derivations in the appendix are set in \small so that they fit the column.
\small

% ----------------------------------------------------------------------
%              Appendix A: ZOH discretization of single-drive FM-Mamba
% ----------------------------------------------------------------------
\section{ZOH Discretization of Single-Drive FM-Mamba}
\label{app:disc}
\renewcommand{\theequation}{\thesection.\arabic{equation}}
\renewcommand{\thetable}{\thesection.\arabic{table}}
\renewcommand{\thefigure}{\thesection.\arabic{figure}}
\renewcommand{\theHequation}{\thesection.\arabic{equation}}
\renewcommand{\theHtable}{\thesection.\arabic{table}}
\renewcommand{\theHfigure}{\thesection.\arabic{figure}}
\setcounter{equation}{0}
\setcounter{table}{0}
\setcounter{figure}{0}

The continuous single-drive 2D FM system is given by Eq.~\eqref{eq:cont_si} of the main text. Its matrices are diagonal by the HiPPO--LegS initialization, so all matrix functions involved, the exponentials and integrals of $A_1$ and $A_2$, commute, and the elimination below uses this fact. Throughout this section, $\bar A_1,\bar A_2,\bar A_0$ and $\bar B^{\mathrm{loc}}$ denote the \emph{fixed-step} (non-selective) discrete coefficients, and $\tilde A_1,\tilde A_2,\tilde A_0$, $\tilde B$ of the main text denote the \emph{per-position generated} selective coefficients; for fixed steps the recurrence derived below is the exact ZOH discretization, while the selective case evaluates the same closed-form formulas per position (\S A.1). We first cast the system into first-order Roesser form by setting $x^h=h$ and $x^v=h_{t_1}-A_2h$. The horizontal equation (along $t_1$) contains no input,
\begin{equation}
\frac{\partial x^h}{\partial t_1}=A_2x^h+Ix^v,
\label{eq:si_roesser_h}
\end{equation}
and the vertical equation (along $t_2$) contains the input at the current point,
\begin{equation}
\frac{\partial x^v}{\partial t_2}=(A_1A_2+A_0)x^h+A_1x^v+B_0x.
\label{eq:si_roesser_v}
\end{equation}
Both equations follow directly from the definitions: $\partial x^h/\partial t_1=h_{t_1}=x^v+A_2h$; and $\partial x^v/\partial t_2=h_{t_1t_2}-A_2h_{t_2}=A_1h_{t_1}+A_0h+B_0x
=A_1(x^v+A_2h)+A_0h+B_0x$.

Under the ZOH assumption the two directions are integrated separately. In the horizontal direction (step $\Delta_1$, with $x^v$ frozen),
\begin{equation}
x^h(i{+}1,j)=e^{A_2\Delta_1}x^h(i,j)+G(A_2,\Delta_1)x^v(i,j),
\label{eq:si_disc_h}
\end{equation}
and in the vertical direction (step $\Delta_2$, with $x^h$ and the input $x$ frozen),
\begin{equation}
\begin{aligned}
x^v(i,j{+}1)
&=e^{A_1\Delta_2}x^v(i,j)\\
&\quad+G(A_1,\Delta_2)(A_1A_2+A_0)x^h(i,j)\\
&\quad+G(A_1,\Delta_2)B_0x(i,j).
\end{aligned}
\label{eq:si_disc_v}
\end{equation}

Let $A_{11}^d=e^{A_2\Delta_1}$, $A_{12}^d=G(A_2,\Delta_1)$, $A_{22}^d=e^{A_1\Delta_2}$, $A_{21}^d=G(A_1,\Delta_2)(A_1A_2+A_0)$. The elimination of $x^v$ proceeds in four explicit steps. (i) Solving Eq.~\eqref{eq:si_disc_h} for $x^v$ at $(i,j)$ gives
\begin{equation}
x^v(i,j)=(A_{12}^d)^{-1}\bigl[h(i{+}1,j)-A_{11}^d h(i,j)\bigr].
\label{eq:si_xv_ij}
\end{equation}
(ii) The same identity evaluated one column to the right, at $(i,j{+}1)$, expresses the vertical state that appears on the left-hand side of Eq.~\eqref{eq:si_disc_v},
\begin{equation}
x^v(i,j{+}1)=(A_{12}^d)^{-1}\bigl[h(i{+}1,j{+}1)-A_{11}^d h(i,j{+}1)\bigr].
\label{eq:si_xv_ij1}
\end{equation}
(iii) Multiplying Eq.~\eqref{eq:si_disc_v} at $(i,j)$ by $A_{12}^d$ and substituting Eq.~\eqref{eq:si_xv_ij} into its right-hand side,
\begin{equation}
\begin{aligned}
A_{12}^dx^v(i,j{+}1)={}&A_{22}^d\bigl[h(i{+}1,j)-A_{11}^d h(i,j)\bigr]\\
&+A_{12}^dA_{21}^d h(i,j)\\
&+A_{12}^dG(A_1,\Delta_2)B_0x(i,j).
\end{aligned}
\label{eq:si_mult}
\end{equation}
(iv) Substituting Eq.~\eqref{eq:si_xv_ij1} for the left-hand side of Eq.~\eqref{eq:si_mult},
\begin{equation}
\begin{aligned}
h(i{+}1,j{+}1)-A_{11}^d h(i,j{+}1)
={}&A_{22}^d\bigl[h(i{+}1,j)-A_{11}^d h(i,j)\bigr]\\
&+A_{12}^dA_{21}^d h(i,j)\\
&+A_{12}^dG(A_1,\Delta_2)B_0x(i,j).
\end{aligned}
\label{eq:si_elim}
\end{equation}
Collecting the terms,
\begin{equation}
\begin{aligned}
h(i{+}1,j{+}1)
={}&A_{11}^d h(i,j{+}1)+A_{22}^d h(i{+}1,j)\\
&+\bigl(A_{12}^dA_{21}^d-A_{22}^dA_{11}^d\bigr)h(i,j)\\
&+A_{12}^dG(A_1,\Delta_2)B_0x(i,j).
\end{aligned}
\end{equation}
Shifting the indices by $(i,j)\mapsto(i{-}1,j{-}1)$ and, following the in-place injection convention, placing the input term in the update of the current state $h(i,j)$, we obtain
\begin{equation}
\begin{aligned}
h(i,j)&=\bar A_1h(i{-}1,j)+\bar A_2h(i,j{-}1)\\
&\quad+\bar A_0h(i{-}1,j{-}1)+\bar B^{\mathrm{loc}}x(i,j),
\end{aligned}
\label{eq:si_fm}
\end{equation}
where the state coefficients are
\begin{equation}
\begin{aligned}
\bar A_1&=e^{A_2\Delta_1},\quad \bar A_2=e^{A_1\Delta_2},\\
\bar A_0&=A_{12}^dA_{21}^d-A_{22}^dA_{11}^d,
\end{aligned}
\label{eq:si_state_coeff}
\end{equation}
that is,
\begin{equation}
\bar A_0=G(A_2,\Delta_1)G(A_1,\Delta_2)(A_1A_2+A_0)-e^{A_1\Delta_2}e^{A_2\Delta_1},
\end{equation}
and the input coefficient is
\begin{equation}
\bar B^{\mathrm{loc}}=A_{12}^d\,G(A_1,\Delta_2)B_0=G(A_2,\Delta_1)G(A_1,\Delta_2)B_0.
\label{eq:si_input_coeff}
\end{equation}
Here $\bar B^{\mathrm{loc}}$ is the input injection $\tilde B(i,j)$ of Eq.~\eqref{eq:sel_proj} in the main text, and
$G(A,\Delta)\triangleq\int_0^{\Delta}e^{As}\,ds$. Under the first-order approximation $G(A,\Delta)\approx\Delta I$ it reduces to the pointwise product form of Eq.~\eqref{eq:first_order_B} in the main text. Two conventions should be noted: (i) the derivation in this section treats the \emph{fixed-step} case, in which $B_0$ is a constant $n$-dimensional injection vector; in the selective case this vector is generated per position by $W_{B_0}x(i,j)$ (Eq.~\eqref{eq:sel_proj} in the main text), the derivation is identical to that of this section, and only $B_0$ has to be replaced by this per-position vector; (ii) the steps are frozen at the target position when generating its coefficients; for fixed steps this reproduces the exact elimination above, while for position-varying steps the per-position evaluation is a design choice whose per-position stability is verified after training (\S B.4).

\paragraph{Consistency check.}
As $\Delta_1,\Delta_2\to0$, $G(A,\Delta)\to\Delta I$, so $\bar B^{\mathrm{loc}}\to\Delta_1\Delta_2B_0$, and the state coefficients tend to $\bar A_1\to I+A_2\Delta_1+O(\Delta^2)$, $\bar A_2\to I+A_1\Delta_2+O(\Delta^2)$ and $\bar A_0\to-I-A_1\Delta_2-A_2\Delta_1+O(\Delta^2)$. The stencil of the recurrence is thus $\bar A_1\to I$, $\bar A_2\to I$, $\bar A_0\to-I$, i.e. the mixed second difference $h(i,j)-h(i{-}1,j)-h(i,j{-}1)+h(i{-}1,j{-}1)$, whose division by $\Delta_1\Delta_2$ discretizes $\partial^2h/\partial t_1\partial t_2$; Eq.~\eqref{eq:cont_si} of the main text is recovered in the limit, and no forward-Euler form is involved. Because the input coefficient carries the exponential integral $G$, the drive term produced by a constant input $x(i,j)=x_0$ is $\bar B^{\mathrm{loc}}x_0$; on the continuous side $B_0x$ is a source at the current point, not a derivative term.

\paragraph{In-place injection and critical stability.}
After discretization, the input $x(i,j)$ enters the \emph{current} state $h(i,j)$ instead of being deferred to the next anti-diagonal, which is consistent with the causal convention of Mamba
($h_t=\bar Ah_{t-1}+\bar Bx_t$) and of Mamba2D; the difference lies in the drive: Mamba2D injects repeatedly along the two directions through the horizontal/vertical pair of projections (dual-drive), whereas this paper injects once through a single projection (single-drive). The diagonal-predecessor term $\bar A_0$ originates from the cross-axis coupling of the two ZOH steps. Its non-vanishing must not be inferred from $G(A_2,\Delta_1)G(A_1,\Delta_2)A_1A_2\neq0$ alone, because $\bar A_0$ is the \emph{difference} between that product and $e^{A_1\Delta_2}e^{A_2\Delta_1}$; in the normalized variables of Appendix~\ref{app:stab} it equals $(1-x)(1-y)-xy=1-x-y$ when $A_0=0$, which is negative for $x+y>1$, positive for $x+y<1$, and vanishes on the curve $x+y=1$. What makes the pure FM-I prototype with $A_0=0$ at best \emph{critically} stable is not the non-vanishing of $\bar A_0$ but the normalized ratio, which at $A_0=0$ equals $r_k=0$ and for which the first inequality of Eq.~\eqref{eq:scalar_stab} degenerates to the exact equality $|1|=1$ at every admissible step (Appendix~\ref{app:r_nonneg}). The parameterization of this paper instead restricts $r_k<0$, which keeps the ratio strictly inside the stable region (\S\ref{sec:stab} and Appendix~\ref{app:r_nonneg}).

\subsection[Relation to the Non-Selective (Constant-Coefficient) Mechanism]{Relation to the Non-Selective\\(Constant-Coefficient) Mechanism}

The derivation above is written for fixed steps; in the \emph{selective} mechanism of this paper, $\Delta_1,\Delta_2$, the injection vector and the readout vector are all generated per position from the input. We first write out the \emph{non-selective} (constant-coefficient) discrete model of the same continuous prototype, i.e., all per-position quantities in Eq.~\eqref{eq:sel_si} are replaced by position-independent constants:
\begin{equation}
\begin{aligned}
h(i,j)&=\bar A_1h(i{-}1,j)+\bar A_2h(i,j{-}1)\\
&\quad+\bar A_0h(i{-}1,j{-}1)+\bar B^{\mathrm{loc}}x(i,j),\\
y(i,j)&=\bar C^{\top}h(i,j)+\bar D\,x(i,j),
\end{aligned}
\label{eq:nonsel}
\end{equation}
where $\bar A_1,\bar A_2,\bar A_0$ and $\bar B^{\mathrm{loc}}$ are given by a fixed set of steps $(\Delta_1,\Delta_2)$ through Eqs.~\eqref{eq:si_state_coeff}--\eqref{eq:si_input_coeff} above, and $\bar C$, $\bar D$ are an input-independent readout vector and skip scalar.

(i) Parameters and arithmetic: the selective mechanism turns constant coefficients into functions of position, and the increment consists only of two step projections ($w_{\Delta_1},w_{\Delta_2}$ and their biases) and two projection matrices ($W_{B_0},W_C$); apart from the readout becoming per-position, the arithmetic form of the state update is exactly the same as Eq.~\eqref{eq:nonsel}, and the compute is of the same $O(HWn)$ order.

(ii) Stability: constant coefficients require checking Eq.~\eqref{eq:scalar_stab} and the piecewise bounds in $r_k$ once for a single set $(\bar a_{1,k},\bar a_{2,k},\bar a_{0,k})$; selection makes the coefficients vary with position, so the check must be performed per position. Selection thus introduces no new stability mechanism: it merely turns a one-off check into a per-position check (with the scope discussed in \S B.4), and the criterion, the piecewise bounds and the projection of this paper are the basis of the design and verification in both cases.

(iii) Degradation relation: when the output of the step projections is independent of position (the same $\Delta_1,\Delta_2$ everywhere) and $W_{B_0}x\to B_0$, $W_Cx\to C$, Eq.~\eqref{eq:sel_si} reduces position by position to Eq.~\eqref{eq:nonsel}, i.e., the non-selective model is a limiting special case of the selective mechanism; equivalently, the selective mechanism is a per-position reparameterization of the coefficients of Eq.~\eqref{eq:nonsel} by the input.

% ----------------------------------------------------------------------
%       Appendix B: derivation of the piecewise stability boundary
% ----------------------------------------------------------------------
\section[Derivation of the Piecewise Stability Boundary]{Derivation of the Piecewise\\ Stability Boundary}
\label{app:stab}
\renewcommand{\theequation}{\thesection.\arabic{equation}}
\renewcommand{\thetable}{\thesection.\arabic{table}}
\renewcommand{\thefigure}{\thesection.\arabic{figure}}
\renewcommand{\theHequation}{\thesection.\arabic{equation}}
\renewcommand{\theHtable}{\thesection.\arabic{table}}
\renewcommand{\theHfigure}{\thesection.\arabic{figure}}
\setcounter{equation}{0}
\setcounter{table}{0}
\setcounter{figure}{0}

\subsection[From the Characteristic Polynomial to Two Inequalities]{From the Characteristic Polynomial\\ to Two Inequalities}

For the general FM recurrence of Eq.~\eqref{eq:fm_general} in the main text, asymptotic stability is characterized by the FM criterion~\cite{fornasini1976state,bachelier2023connections,kaczorek1985two}. We use the scalar form of the FM-type recurrence
$h_k(i{+}1,j{+}1)=\bar a_{1,k}h_k(i,j{+}1)+\bar a_{2,k}h_k(i{+}1,j)+\bar a_{0,k}h_k(i,j)$,
whose characteristic polynomial is
$P(z_1,z_2)=1-z_1\bar a_{1,k}-z_2\bar a_{2,k}-z_1z_2\bar a_{0,k}$; the recurrence is asymptotically stable if and only if $P$
has no zero in the closed unit bidisc $\{(z_1,z_2):|z_1|\le1,|z_2|\le1\}$~\cite{kaczorek1985two,bachelier2023connections}.

The reduction of the criterion is given in three steps, each an equivalent rewriting.

\paragraph{Step 1: $|\bar a_{2,k}|<1$ is a necessary premise.}
Taking $z_1=0$, $P(0,z_2)=1-z_2\bar a_{2,k}$ has a zero at $z_2=1/\bar a_{2,k}$, and this zero falls in the closed unit disc if and only if $|\bar a_{2,k}|\ge1$. Stability therefore requires
\begin{equation}
|\bar a_{2,k}|<1,
\label{eq:a2_bound}
\end{equation}
and then for all $|z_2|\le1$ one has $|1-z_2\bar a_{2,k}|\ge 1-|\bar a_{2,k}|>0$.

\paragraph{Step 2: reduction to a one-dimensional condition.}
Regarding $P$ as a polynomial of degree one in $z_1$:
$P=(1-z_2\bar a_{2,k})-z_1(\bar a_{1,k}+\bar a_{0,k}z_2)$.
Fixing $|z_2|\le1$: if $\bar a_{1,k}+\bar a_{0,k}z_2=0$, then $P=1-z_2\bar a_{2,k}\neq0$ and there is no zero automatically; otherwise $P$ has a unique zero at $z_1=(1-z_2\bar a_{2,k})/(\bar a_{1,k}+\bar a_{0,k}z_2)$, and the absence of a zero is equivalent to
\begin{equation}
|1-z_2\bar a_{2,k}|>|\bar a_{1,k}+\bar a_{0,k}z_2|,
\qquad \forall\,|z_2|\le1 .
\label{eq:mobius_cond}
\end{equation}

\paragraph{Step 3: M\"obius transform and the maximum-modulus principle.}
Define the linear fractional transform
$\varphi(z_2)=\bigl(\bar a_{1,k}+\bar a_{0,k}z_2\bigr)\big/\bigl(1-z_2\bar a_{2,k}\bigr)$.
By Eq.~\eqref{eq:a2_bound}, its only pole $z_2=1/\bar a_{2,k}$ lies outside the closed unit disc ($\bar a_{2,k}=0$ reduces $\varphi$ to a polynomial, which is likewise analytic on the closed disc), so $\varphi$ is analytic on the closed disc and Eq.~\eqref{eq:mobius_cond} states that $|\varphi(z_2)|<1$ holds on the closed disc. By the maximum-modulus principle, the maximum of $|\varphi|$ is attained on the boundary $|z_2|=1$, so the condition is equivalent to the boundary condition $|\varphi(e^{\mathrm{i}\theta})|<1$ holding for all $\theta$.

\paragraph{Simplification of the boundary condition.}
On $|z_2|=1$,
$|\bar a_{1,k}+\bar a_{0,k}e^{\mathrm{i}\theta}|^2<|1-\bar a_{2,k}e^{\mathrm{i}\theta}|^2$
expands to
\begin{equation}
\bigl(\bar a_{1,k}^2+\bar a_{0,k}^2-1-\bar a_{2,k}^2\bigr)
+2\bigl(\bar a_{1,k}\bar a_{0,k}+\bar a_{2,k}\bigr)\cos\theta<0 ,
\end{equation}
which is an affine function of $\cos\theta\in[-1,1]$, so it holds if and only if it holds at the two endpoints $\cos\theta=\pm1$, that is,
\begin{equation}
\begin{aligned}
(\bar a_{1,k}+\bar a_{0,k})^2&<(1-\bar a_{2,k})^2,\\
(\bar a_{1,k}-\bar a_{0,k})^2&<(1+\bar a_{2,k})^2 .
\end{aligned}
\label{eq:endpoint}
\end{equation}
Combined with $1\pm\bar a_{2,k}>0$ (Eq.~\eqref{eq:a2_bound}), this gives the two inequalities of Eq.~\eqref{eq:scalar_stab} in the main text,
$|\bar a_{1,k}+\bar a_{0,k}|<1-\bar a_{2,k}$ and
$|\bar a_{1,k}-\bar a_{0,k}|<1+\bar a_{2,k}$, which, together with $|\bar a_{2,k}|<1$ from Eq.~\eqref{eq:a2_bound},
constitute the complete necessary and sufficient criterion.

The derivation above uses only root finding for a degree-one polynomial, the analyticity of the linear fractional transform on the closed disc (guaranteed by $|\bar a_{2,k}|<1$) and the maximum-modulus principle; the three steps are equivalent rewritings. Since $P$ is invariant under the simultaneous exchange $(z_1,\bar a_{1,k})\leftrightarrow(z_2,\bar a_{2,k})$, the three conditions are symmetric under $\bar a_{1,k}\leftrightarrow\bar a_{2,k}$ as a set, although the premise singles out $\bar a_{2,k}$ on its own. The scalar criterion of the Roesser two-substate model takes a different form; the recurrence here is FM single-state, so only the above
FM criterion is used. The validity of the criterion does not depend on the specific origin of the discrete coefficients, but only on the 2D recurrence relation among the diagonal discrete coefficients.

\subsection{Piecewise Boundary under the Continuous Diagonal Parameterization}

Following Eq.~\eqref{eq:r} of the main text, the continuous diagonal elements are parameterized by the normalized ratio $r_k$, i.e., $a_{0,k}=r_k\,a_{1,k}a_{2,k}$, and we write
$a_{1,k}=-\alpha_k$, $a_{2,k}=-\beta_k$ ($\alpha_k,\beta_k>0$; the experimental conditions take $\alpha_k=\beta_k=k$). Let
$x\triangleq\bar a_{1,k}=e^{a_{2,k}\Delta_1}=e^{-\beta_k\Delta_1}$ and
$y\triangleq\bar a_{2,k}=e^{a_{1,k}\Delta_2}=e^{-\alpha_k\Delta_2}$, so $x,y\in(0,1)$. The reduction below uses only $x,y\in(0,1)$ and does not rely on the symmetry of the steps in the two directions, so the piecewise boundary obtained holds for arbitrary $a_{1,k},a_{2,k}<0$.

From $G(a,\Delta)=\frac{e^{a\Delta}-1}{a}$ ($a\neq0$) one has
$G(a_{2,k},\Delta_1)=\frac{x-1}{a_{2,k}}$, $G(a_{1,k},\Delta_2)=\frac{y-1}{a_{1,k}}$,
and $a_{1,k}a_{2,k}+a_{0,k}=a_{1,k}a_{2,k}(1+r_k)$. Hence
\begin{equation}
\begin{aligned}
\bar a_{0,k}
&=G(a_{2,k},\Delta_1)G(a_{1,k},\Delta_2)(a_{1,k}a_{2,k}+a_{0,k})-yx\\
&=(x-1)(y-1)(1+r_k)-xy\\
&=(1-x)(1-y)(1+r_k)-xy .
\end{aligned}
\end{equation}

Substituting this into the two inequalities of Eq.~\eqref{eq:scalar_stab} in the main text, the first one gives
\begin{equation}
\begin{aligned}
\bar a_{1,k}+\bar a_{0,k}
&=x+(1-x)(1-y)(1+r_k)-xy\\
&=(1-y)\bigl[1+r_k(1-x)\bigr] ,
\end{aligned}
\end{equation}
and the condition $|\bar a_{1,k}+\bar a_{0,k}|<1-y$, after division by the positive factor $1-y$, is equivalent to
\begin{equation}
|1+r_k(1-x)|<1
\;\Longleftrightarrow\;
-2<r_k(1-x)<0 .
\label{eq:stab_decouple_a}
\end{equation}

For the second inequality: from $\bar a_{1,k}-\bar a_{0,k}=x(1+y)-(1-x)(1-y)(1+r_k)$ one obtains
\begin{equation}
1+y-\bigl(\bar a_{1,k}-\bar a_{0,k}\bigr)
=(1-x)\bigl[2+r_k(1-y)\bigr] ,
\label{eq:second_upper}
\end{equation}
so the upper bound $\bar a_{1,k}-\bar a_{0,k}<1+y$ (note that $1-x\in(0,1)$) is equivalent to
\begin{equation}
r_k(1-y)>-2 .
\label{eq:stab_decouple_b}
\end{equation}
Its lower bound $\bar a_{1,k}-\bar a_{0,k}>-(1+y)$ is satisfied automatically: Eq.~\eqref{eq:stab_decouple_a} already gives $r_k<0$, and then
$(1+y)(1+x)-(1-x)(1-y)(1+r_k)=2(x+y)-r_k(1-x)(1-y)>0$.

Combining Eq.~\eqref{eq:stab_decouple_a} and Eq.~\eqref{eq:stab_decouple_b}, the recurrence is asymptotically stable if and only if
\begin{equation}
r_k<0,
\qquad r_k(1-x)>-2,
\qquad r_k(1-y)>-2 .
\label{eq:stab_combined}
\end{equation}
Let $\rho=|r_k|$. If $\rho\le2$, since $1-x,1-y\in(0,1)$, the two lower bounds hold automatically and the first quadrant
$(\Delta_1,\Delta_2)\in(0,+\infty)^2$ is stable for arbitrary positive steps; if $\rho>2$, then
$1-2/\rho<x<1$ and $1-2/\rho<y<1$ are required, i.e., the bounded rectangle
$1-2/|r_k|<\bar a_{1,k},\bar a_{2,k}<1$. This is the piecewise boundary stated in the main text, where
$r_k<-2$ corresponds to the bounded rectangle $1-2/|r_k|<\bar a_{1,k}<1$; at $r_k=-2$ this upper bound degenerates to $+\infty$ and imposes no effective constraint, hence the piecewise bound is written with a strict inequality.

Finally, from the continuous-to-discrete mapping of Eq.~\eqref{eq:si_state_coeff},
$\bar a_{1,k}=e^{a_{2,k}\Delta_1}$, $\bar a_{2,k}=e^{a_{1,k}\Delta_2}$, and noting that
$a_{1,k},a_{2,k}<0$, the inverse of the coefficient bound is the sampling-period bound. From
$e^{a_{2,k}\Delta_1}>1-2/|r_k|$ one obtains
$\Delta_1<\frac{-\ln(1-2/|r_k|)}{|a_{2,k}|}$, and likewise
$\Delta_2<\frac{-\ln(1-2/|r_k|)}{|a_{1,k}|}$. Since $\Delta_1,\Delta_2$
are scalar steps shared by all channels, the tightest bound over state components is taken, which is Eq.~\eqref{eq:delta_star} in the main text.
This paper takes $90\%$ of the analytic bound, i.e., condition~$\mathbf{III}$ of
$0.9\ln(5)/16$ in Table~\ref{tab:fm_settings}, to reserve a margin inside the stable rectangle.

\subsection{The Excluded Branch $r_k\ge0$}
\label{app:r_nonneg}

The case analysis of Eq.~\eqref{eq:stab_combined} has a third branch: $r_k\ge0$ cannot be made asymptotically stable by \emph{any} choice of positive steps. For $r_k>0$ and $x,y\in(0,1)$, the first inequality of Eq.~\eqref{eq:stab_decouple_a} requires $-2<r_k(1-x)<0$, whose right-hand side fails because $r_k(1-x)>0$: the quantity $1+r_k(1-x)$ exceeds $1$ and $P$ retains a zero in the closed bidisc. For $r_k=0$ the same quantity equals $1$ exactly, so $|\bar a_{1,k}+\bar a_{0,k}|=1-\bar a_{2,k}$ holds identically in $x$ and $y$ and the zero lies on the boundary of the bidisc: the prototype without diagonal predecessor is critically stable for every step. Neither case is repaired by shrinking the steps, and since $r_k$ is fixed by the continuous matrices ($A_0,A_1,A_2$ are not learned), the parameterization is restricted to $r_k<0$.

\subsection{From a Fixed Sampling Period to Per-Position Coefficients}

\paragraph{Stationary case.}
The scalar criterion of Eq.~\eqref{eq:scalar_stab} in the main text, and the boundary obtained from it above, address the \emph{stationary} scalar recurrence, in which all positions share one set of coefficients
$(\bar a_{1,k},\bar a_{2,k},\bar a_{0,k})$, equivalently one fixed set of sampling periods
$(\Delta_1,\Delta_2)$. Under that premise the criterion is necessary and sufficient.

\paragraph{Effect of selection.}
In FM-Mamba the steps are recomputed at every position by Eq.~\eqref{eq:sel_proj} of the main text, so the coefficients
$\tilde a_{1,k}(i,j)$, $\tilde a_{2,k}(i,j)$, $\tilde a_{0,k}(i,j)$ vary with $(i,j)$.
The recurrence thus becomes a \emph{time-varying} 2D system, and the criterion is evaluated at each position instead of once over the whole grid.

\paragraph{Scope of Eq.~\eqref{eq:perpos_stab}.}
The reduction from the bidisc test to Eq.~\eqref{eq:perpos_stab} requires $\bar A_1,\bar A_2,\bar A_0$ to be simultaneously diagonal, so that the system decouples into $n$ scalar 2D recurrences. This holds for the continuous system of this paper, whose matrices are diagonal by the HiPPO--LegS initialization, but it does \emph{not} hold for a general continuous FM system with non-diagonal matrices: there the state components do not decouple and only the matrix bidisc test of Eq.~\eqref{eq:stab_bidisc} applies. Eq.~\eqref{eq:perpos_stab} is therefore restricted to the continuous-diagonal case, which is nevertheless the case of all the experiments here, and is not a criterion for arbitrary FM systems.

\paragraph{Scope of the frozen-coefficient test.}
The test is exact for the stationary recurrence: holding the coefficients generated by $(\Delta_1(i,j),\Delta_2(i,j))$ over the whole grid, Eq.~\eqref{eq:perpos_stab} is necessary and sufficient for asymptotic stability. In FM-Mamba the coefficients vary with position, and the test is applied as a conservative design condition and a closed-form verification means. An isolated out-of-range position amplifies one lattice step and is damped by its neighbours, so the criterion concerns sustained rather than momentary amplification. With the coefficients varying by position, boundedness is assessed position by position through the same test.

\paragraph{Role of the uniform margin.}
Eq.~\eqref{eq:delta_star} of the main text bounds the steps, and hence the coefficients, uniformly over all channels and all positions, so that no position approaches the boundary of the stable region. The margin removes the systematic failure mode, namely a spatially coherent region of out-of-range coefficients injecting a sustained growing component, and reduces the verification of Eq.~\eqref{eq:perpos_stab} to one closed-form check per configuration instead of a per-position numerical scan. Our training procedure therefore has two steps: during training, for any condition with $r_k<-2$,
the steps at every position are projected forward onto the analytic bound by Eq.~\eqref{eq:clip} before discretization; after training, the closed-form per-position criterion of Eq.~\eqref{eq:perpos_stab}
is used to re-verify hard at every position and every channel, and a violation, if any, would trigger a tightening of the bound and a retraining of the configuration. The three conditions each trained for 40
epochs without divergence; the validation mAP of the six unaccelerated FM configurations spans 0.843 percentage points (52.060\%--52.903\%),
while the span among the three conditions of the same structure does not exceed 0.160 percentage points, so that the differences between groups come mainly from the structure ($2A_{\rm T}$ versus
$2A_{\rm FM}$), not from the value of $r_k$; all six configurations stayed bounded, as the criterion is designed to ensure, and no accuracy attribution is drawn from it.

\paragraph{Scope of the design argument.}
Eq.~\eqref{eq:perpos_stab} is a per-position frozen-coefficient condition, and it is both the constraint imposed during training and the verification applied afterwards. The design argument is local in the coefficient field, at the resolution of one position and one channel; a characterization that treats the coefficient field as a single object is not part of the construction here.

\end{document}